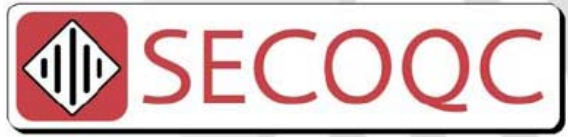

# SECOQC
# BUSINESS WHITE PAPER

## QUANTUM CRYPTOGRAPHY:
### An Innovation in the Domain of
### Secure Information Transmission

Editing author: S. Ghernaouti-Hélie

Contributing authors: S. Ghernaouti-Hélie, I. Tashi; Th. Länger; C. Monyk

September 2008

## 1 ABSTRACT

In contemporary cryptographic systems, secret keys are usually exchanged by means of methods, which suffer from mathematical and technology inherent drawbacks. That could lead to unnoticed complete compromise of cryptographic systems, without a chance of control by its legitimate owners.

Therefore a need for innovative solutions exists when truly and reliably secure transmission of secrets is required for dealing with critical data and applications. **Quantum Cryptography** (QC), in particular **Quantum Key Distribution** (QKD) can answer that need.

The business white paper (BWP) summarizes how secret key establishment and distribution problems can be solved by quantum cryptography. It deals with several considerations related to how the quantum cryptography innovation could contribute to provide business effectiveness. It addresses advantages and also limitations of quantum cryptography, proposes a scenario case study, and invokes standardization related issues. In addition, it answers most frequently asked questions about quantum cryptography.

The aims of this white paper are to:

- **explain** the innovation produced by the SECOQC consortium related to quantum cryptography for secure information transmission;

- **promote** the use of quantum cryptography by describing the business advantages of integrating such mechanisms and quantum networks in existing telecommunication architecture;

- **facilitate** decision making processes for adoption of quantum cryptography solutions by business managers for the benefit of public or private institutions;

- **raise** public awareness of the possible use of the application of quantum cryptography to enhance actual cryptographic security mechanisms.

**SECOQC Business White Paper addresses quantum cryptography business related issues. Consequently, it does not give mathematical or physical details of classical or quantum cryptography. The reader does not need to have a mathematical, physical, computer science or telecommunication background**.

## 2 EXECUTIVE SUMMARY

Offering a good level of data **confidentiality** is nowadays becoming an important feature for **business realisation** and **competitiveness**. SECOQC provides a tool based on quantum cryptography that will enable organisations to exchange critical information with guaranteed confidentiality, knowing that their vital **informational assets are secure from industrial espionage and other forms of illegal activities**.

Quantum cryptography provides **long-term security** and thus **conforms** to the requirements of a number of recent **legal regulations** for protecting information.

For any institution, as for examples financial entities, governments, military agencies, airports or critical infrastructures as power stations, etc., the necessity for secure communication takes **highest priority** when the exchanged information is considered as **critical**. For such institutions a security breach in the transmitted information could be a disaster that could not be covered by insurance mechanisms.

Contemporary commercial cryptographic solutions are not provably secure, because of their **intrinsic vulnerabilities**. Thus, the main component of secure transmission of confidential data, the cryptographic key, cannot be considered as totally secure. Methods and systems considered as acceptably secure today have a **significant risk** of becoming weak tomorrow.

Since the recent past, quantum cryptography has only been carried out in experimental projects. Only physical scientists were interested in quantum information theory. Nowadays this is evolving and **more and more information security specialists focus on the practical utilisation of quantum cryptography** to increase the security level of confidential data transmission.

Quantum cryptography is no more a subject reserved to scientific conferences for physics experts; it is becoming an **alternative solution** meeting the real needs of **demanding users**. Innovation in optical technologies makes it possible by providing the market with devices capable of generating, detecting and guiding single photons; devices that are affordable within a commercial environment. They can be integrated in **existing telecommunication architectures** to enforce data communication security. As demonstrate by SECOQC, it is possible to use them to design point-to-point links, virtual private networks or telecommunication networks.

Quantum cryptography offers effective cryptographic mechanisms without passing by a third party or a certification authority as requested when using a public key infrastructure (PKI). In existing highest-security cryptography systems, cryptographic keys have to be distributed beforehand, which is a weakness of these system. Quantum cryptography brings an **efficient solution** to generate, manage and transmit the cryptographic keys in a very secure way.

**Quantum key establishment and distribution processes** reach the **highest level of security guarantee** because they rely upon **laws of physics** independent from the time variable and will never be threatened by increased computing power or by progress made in mathematical discipline.

In addition, quantum technology is now available to manufacture **true random number generators**. This is of the greatest interest to avoid predictable and deterministic computer behaviour and prevent several types of attacks against cryptographic systems. Quantum cryptography has been developed to be secure against a wide range of attacks; whatever is the computer power a malevolent can rely upon.

Any organisation using quantum cryptography, which is an **inherently secure** way to communicate secrets, would not be anymore at the same level of insecurity as its competitors. Quantum cryptography brings a real **competitive and reputation advantage**.

Rethinking fundamentals in cryptography is the only solution to **develop a new vision of security** for the benefit of transactions that are critical for **institutions and people**.

With quantum cryptography, institutions and people have for the first time the means to be sure that their data are under their **own control** and cannot be obtained by eavesdroppers without knowledge of sender or recipient.

Today, quantum cryptography solutions are available to build **secure communication infrastructures.**

**CONTENTS**

## 3 THE EUROPEAN PROJECT SECOQC

Facing the challenge of transforming the **European Union** into the most dynamic and competitive knowledge-based economy in the world, the European Commission has procured the sixth framework programme (FP6) fund, available for research, divided into seven European Research Areas (ERA). The integrated project SECOQC fits into the **Information Society Technologies** area. SECOQC, being the acronym for "**Secure Communication based on Quantum Cryptography**", was launched to contribute to evolve quantum cryptography into an instrument that can be operated in an economic competitive environment [1].

To put all the chances on SECOQC's side to reach scientific excellence, improved competitiveness and innovation, the consortium has been composed of different European institutions with several complementary fields of expertise and competences. All in all 42 partners are involved in this European project, and they work hand in hand to propose an effective information security solution. These various backgrounds enable the project and its completeness to be considered from all points of views. A wide range from conceptual and pragmatic issues belonging to quantum physics, to network architecture, to legal issues and to socio-economics impacts of new technology adoption, has been addressed in this project.

SECOQC's aim is to provide companies and public institutions with an urgent need of confidential communication with an efficient instrument that qualitatively improves security levels of information transmission via public channels. Therefore the goal is to design a network for secure long-range communication building upon a Quantum Key Distribution (QKD) enabling technology. For that, the co-operation within SECOQC is divided into two main activity blocks: one covering the quantum physical devices and the other the design and set up of the communication infrastructure (cryptographic protocols, communication network, and system integration, in accordance to user requirements).

Through increased co-operation, greater complementarities and improved co-ordination between relevant actors at all levels of the consortium, great results have been achieved to resolve security and trust problems for quantum cryptography technology. Results have been presented to the international scientific community through publications, international conferences and symposiums. **The first live demonstration of a working quantum key distribution network** will take place in **Vienna** in the framework of the SECOQC Demonstration and International Conference (**Oct. 8-10, 2008**). Many awards, prizes and best papers have been awarded to the researchers for the quality and the relevance of their contributions.

## 4 INFORMATION SECURITY FOR INSTITUTIONS' EFFECTIVENESS AND COMPETITIVENESS

### 4.1 Need for strong information confidentiality

The more valuable information is the more eavesdroppers will be motivated to find out how to reach and exploit it. Even data intrinsically and individually considered as non-critical could become critical when being accurately correlated, in regards of **economical and industrial espionage** related issues. Currently, unauthorised third

...............SECOQC Business White Paper – September 2008

parties routinely and systematically attack communication and data transmission over public networks, and put in danger data confidentiality and integrity.

Electronic surveillance systems, blended with massive computing power render almost any "plain text" or encrypted communication completely vulnerable. While electronic surveillance systems currently do not have the cryptanalytic power to endanger strong encryption, attempts are made to subvert strong cryptography into a weak one by imposing international "key escrow" systems. Even cryptographic procedures that are currently considered as secure are becoming increasingly vulnerable. Major cryptography solutions widely used at the moment could be threatened and will definitely become obsolete in a near future.

### 4.2 Need to encrypt data and to share secret keys

Today optical fibres have replaced almost all copper links in worldwide digital communication networks. Optical fibres are thus often used to carry confidential information despite of intrinsic vulnerabilities. Intercepting information transmitted over an optical fibre is not only possible but also quite easy to do [2, 3]. As any telecommunication is intrinsically vulnerable to eavesdropping, cryptography is routinely used to protect data transmission in order to prevent eavesdroppers from being able to use the information that they capture.

**Cryptographic mechanisms** contribute to offer **data confidentiality**. The word "cryptography" comes from Greek meaning "**secret writing**" - it is the science of concealing meaning. Cryptanalysis is the breaking of codes [4]. Cryptography is a deep mathematical subject, but this paragraph will retain only tow points: the need to use cryptographic keys and algorithms to be able to encrypt data.

Whatever the cryptosystem is, for data communication confidentiality, sender and receiver have to use keys and cryptographic protocols to encrypt or decrypt a plain text [5].

There are two crucial moments to be considered when dealing with the **encryption keys** used to protect **sensitive information**: the **key generation process** and **the key distribution process** [6].

For the moment the **key establishment problem** is a crucial problem that the cryptography community has to face in order to propose **truly secure confidentiality tools over unreliable transmission systems.**

The key generation process needs the generation of some truly random numbers and the key distribution process needs some "secure way" to transport the key between the sender and the receiver.

In order to decrypt intercepted information, an eavesdropper has to obtain the key. As a rational person, he / she will concentrate all his effort either on the key generation process by guessing or deducing the key or on the key distribution process by trying to intercept the key during its exchange between the sender and the receiver.

The strength of a cryptographic system depends on key length, key renewal rate, key entropy and key quality as well as on the cryptographic algorithm robustness [7].

...

The robustness of any cryptographic system largely used over the Internet is based on mathematical procedures. Deep knowledge in mathematics associated with large computing power could in principle give ways to recover cryptographic keys during the keys distribution phase (keys exchange). That will completely compromise the data communication enciphered by such keys. Another issue related to cryptography is the lifetime of the information considered as confidential (secret).

Breaking cryptographic mechanisms depends on the available computing power. It will be less and less difficult to achieve, regarding the growing computation capacities and grid availability. Key exchange is therefore vulnerable to technological progress. It means that even today the scheme remains only relatively secure and there is no guarantee about the future. What is appears mathematically unrealistic today may well be achieved tomorrow.

Several questions arise when using cryptographic mechanisms, among them:

- *Will confidential data encrypted by classical cryptographic mechanisms still remain confidential in the long term?*

- *How long will data considered as secure today stay confidential?*

- *Can encrypted data remain secure being conscious that one can compromise the secret key?*

- *How to generate pure random numbers in order to create non-predictable secret keys?*

At present, only **quantum cryptography** can contribute to answer these critical issues.

The use of quantum cryptography for the distribution of secrets offers a leverage for achieving **optimal security** for the transmission of sensitive data [8]. The quantum security relies on the intrinsic secrecy of cryptographic keys guaranteed by **quantum physics** [9, 10]. This permits to reveal an eavesdropper's presence by observing the amount of perturbation. If the perturbation stays below a certain bound, this is a proof of key confidentiality and integrity.

With security **based on physical laws** rather than on hard algebraic problems, quantum cryptography is of the greatest interest to avoid breaking of the algorithm due to predictable and deterministic computer behaviour. Furthermore, security based on physical assumption is more stable as time passes than security based on computational assumptions [11] and will prevent several types of existing attacks. The stability of laws of physics gives **highest guarantees** to obviate all dangers related to technology evolution. Quantum cryptography can remain secure even in case that extremely powerful quantum computer would be available!

Quantum cryptography allows placing **security in a long run vision**.

Quantum cryptography **complements conventional cryptographic techniques** to raise security of data transmission to an unprecedented level.

### 4.3 Need for efficient tools to guarantee information security and satisfy conformance issues

**Several regulations** like the **Sarbanes-Oxley Act** [12] and the **Basel II agreements** [13] have recently emerged. These regulations are directly or indirectly related to the way information systems are organised and secured within the organization. They share some key concepts such as **accountability**, **protection** of personal private information, disclosure policies and **integrity** of reported information that for example can be linked to information security criteria.

In fact, many regulations define some requirements concerning informational risks without specifying them in a more detailed way [14]. **Regulatory compliance issues** as well as **responsibility** and imputability will in the near future continue to rise in importance and will become more and more demanding regarding **information security assurance** [15].

More often, organisations consider the **need for compliance** as being a **catalyst** in resolving **long-overlooked security** problems but at the same time, the need to satisfy regulatory requirements increases information security management complexity level.

To reduce the complexity of the management task, **managers have to rely upon provable and reliable confidentiality technical tools** [16]. Quantum cryptography can contribute to answer this issue.

A **risk management process** has to consider all risk components in order to choose or to propose the most appropriate countermeasures. Being concerned by the costs that countermeasures generate, security managers have to make a cost-benefit analysis in order to spend limited resources appropriately to get good results [17]. This implies to be pro-active and means to reduce the risks and their impacts by decreasing the number of vulnerabilities. Reducing vulnerabilities is becoming the crucial part of risk management [18].

- *How can threats against cryptographic mechanisms be diminished?*
- *How can vulnerabilities of data confidentiality be decreased?*

These issues can be solved by integrating quantum cryptography into existing security mechanisms [19-24].

Researchers in the field consider quantum cryptography as the only true secure key-distribution technology[25].

### 4.4 Information security is a business enabler

Organisations have to face the following security challenges:

- **to protect** information and communication infrastructures;
- **to prevent** information leakage,
- **to avoid** information and communication technologies misuses or errors

   (intentional or not);

- **to guarantee** business continuity;

- **to keep** security simple and cost effective.

Because information security is a business enabler organizations have to protect the confidentiality of their digital assets in order to:

- **Prevent** direct and indirect losses;

- **Avoid** reputation damage;

- **Ensure** business continuity;

- **Stay** competitive.

## 5 QUANTUM CRYPTOGRAPHY IS AN INNOVATIVE SOLUTION TO CLASSICAL CRYPTOGRAPHY LIMITATIONS

### 5.1 Quantum cryptography and quantum key distribution (QKD)

Quantum cryptography, or, more exactly, **quantum key distribution** (QKD) is the science that studies how to generate perfect random keys between two parties which are connected by a quantum channel [26]. The term quantum cryptography should be understood as quantum key distribution through this document.

Quantum cryptography was not invented as solution to an urgent demand; rather it originated from pretty theoretical speculations on the power that is added to information theory by using quantum mechanical systems. Quite surprisingly, in the course of years, it turned into an increasingly interesting application of quantum mechanics that can have practical advantages for **information technology** [11].

Quantum cryptography uses photons to transmit valuable information. According to Heisenberg's Uncertainty Principle (1927)[1], it is not possible to observe a quantum object without modifying it. Linked to the very quantum physical characteristics of a photon, these properties are used to verify and certify that the information-photon has not been eavesdropped upon or intercepted [9, 10]. The photons are subsequently used to generate and exchange secrets (keys) between two remote sites through an optical fibre link, confirming thus their secrecy.

In telecommunication networks, light is routinely used to exchange information, but encoding the value of a digital bit on a single light photon to create a secret key then used to encrypt sensitive data to be exchanged is an **innovation** in generating and distributing secrets [27]. For each bit of information, pulses containing a single photon are emitted and sent through an optical fibre to the receiver. The deciphering task is impossible for any eavesdropper because if the eavesdropper wants to obtain the value of a bit he or she must observe the photon. By doing so, the eavesdropper will disturb the quantum system and reveal his or her presence.

---

[1] http://www.aip.org/history/heisenberg/p08.htm

One of the crucial features of the key, the length, is totally flexible in the case of Quantum Key Distribution (QKD). This flexibility is also the case in classical cryptography, but the advantage is that by the use of QKD, the time needed for key generation is not affected by exponential increase as it is for longer asymmetrical keys. This advantage relies on the fact that QKD is a point-to-point primitive using a symmetric cryptographic key distribution mechanism with highest security level. With quantum cryptography the objective of using real random numbers and random processes is reached. This increases the transmission security level and obviates the need to increase the complexity of the key – it is provably secure against any attack irrespective of available computing power or the availability of any other resources.

 "Quantum cryptography can benefit from the perfect secrecy offered by the One-Time-Pad function and from the fact that the keys established by QKD are unconditionally secure, the messages exchanged over such unconditionally secure links enjoy one security property that can be called "everlasting secrecy" [10].

Quantum cryptography has the potential to complement conventional cryptographic techniques to raise the security of data transmission fibre links to an unprecedented level. It is considered as one of the ten technologies that will change the world.[2]

Quantum cryptography security is always relying on an implicit assumption that the persons who want to communicate with each other must be located inside safe environments and belong to a trusted circle.

Quantum cryptography technology has not been intensively tested or attacked and in the same time validated by a large number of users. A strong historical security is rather difficult to claim since very few organizations have yet implemented quantum cryptography in their security systems

## 5.2    A case study example

Luxembourg is a small country, which does not exceed 100 km of length and 57 km of width. However there are over 200 banks, 1785 collective investments in transferable securities (UCITS) and more than 142 other financial related organisations[3].

The service sector is the most important sector within Luxembourg's economy. Something like 100 billion of dollars is the net benefit of the financial market of Luxembourg. According to its importance within the worldwide financial market, Luxembourg is focusing a lot on the inter banks transactions market which is the fourth in size in Europe. On the other hand Luxembourg is considered as a very privileged place for private banking. This is based principally on informational assets and on data and transaction confidentiality. Relying upon strong confidentiality mechanisms to protect secrets is a major concern for all economic actors of Luxembourg's financial place and for the overall economy of the country. Recent details of significant security breaches in every day financial business, where millions of dollars have been lost, reinforce the need for more robust security to protect information on transmission. Considering all these characteristics of such an important financial market, an inherently secure communication network is required. Luxembourg as financial place could benefit from operating quantum security. Due to the limited specific geographic environment of Luxembourg, distance limitation of quantum key distribution should not be any more a concern.

---

[2] MIT technology Review, February 2003 and Newsweek, July 7 2003
[3] Source http://www.assemblee-nationale.fr/11/rap-info/i2311-51.asp

Banks can actually use quantum cryptography in the following types of transaction: bank building to another bank building of the same bank company or/and cash dispenser to the bank, assuming the fact that the distance connecting the remote points is less than 100 km. So, the use of quantum cryptography based on optical fibre is possible.

In this case, either a big amount of data could be exchanged or a tiny amount of data that could be transmitted frequently between systems. Transmission from cash dispenser to bank (if the cash dispenser is not in the bank itself) can also be done using quantum cryptography based on optical fibre to avoid the danger that a malicious person could intercept the communication between the bank and the cash dispenser and modify it (e.g. credit some bank account or change the identification of the debited account).

Integrity and confidentiality of the transmitted data could then be ensured by the use of transactions secured with quantum cryptography.

Because quantum cryptography allows securing encrypted links, it can be used also to design highly secure virtual private networks (VPN) [22].

### 5.3 A real quantum cryptographic networks: SECOQC QKD Network

The presentation of the SECOQC quantum-cryptographic network marks the successful termination of the project after four and a half years (Vienna October 8th, 2008).

Several typical applications, like telephony and video conferencing are secured with cryptographic keys distributed over a QKD network with six nodes and eight links.
The node modules are provided by their developer, Austrian Research Center in Vienna, and, the optical fibre infrastructure is provided by Siemens Austria. QKD Links come from five different technologies: Plug-and-Play system by idQuantique from Geneve/Switzerland, Coherent One-Way system from GAP Optique, Geneva, Switzerland and Austrian Research Center, Continuous-Variables system from Centre national de la Recherche Scientifique (CNRS) and THALES Research and Technology, both from Paris, France, One Way Weak Pulse System from Toshiba Research of UK, as well as an Entangled Photons system from University Vienna and Austrian Research Centers. The logical topology of the system is a square with diagonals plus one further long link. Average distance between the nodes is between 20 and 30 kilometres, with the longest link of 83 kilometres [28].

The network is of the trusted repeater type with a special emphasis on the development of the network architecture and the corresponding protocols [29]. The **SECOQC network** is solely used for the creation of secure Keys and their distribution to the intended receivers. The so-called '**Network of Secrets**' is organised in three layers, being QKD link layer, transport layer, and network layer.

Key creation and distribution is completely hidden in the lowest layer. A distributed management and routing system provides availability through redundancy of paths through the network, as well as traffic and congestion control. With so called 'secret sharing', i.e. the selection of disjunctive paths for different portions of the cryptographic keys, with recombination of the final key only at its final destination, trust assumptions on single nodes of the trusted repeater network can even be relaxed.. A central design feature of the SECOQC Network of Secrets is that cryptographic keys are stored and managed in dedicated key stores that are implemented in the so-called node module of the network node, and not in the single point-to-point links, which enables the keys on a network level. This node module also provides the public channel for all nodes originating (or ending) in the node, and its cryptographic functionality, like authentication and encryption on request [28].

## 6 USING QUANTUM CRYPTOGRAPHY: SOME ADVANTAGES AND LIMITS

### 6.1 Building confidence

Organizations that offer a reliable security level will have easiness to build a trusty relationship with its stakeholders, partners or clients. Quantum cryptography will contribute to develop confidence and to raise practical market related advantages.

### 6.2 Being in regulatory conformity

Several regulations have emerged during the last years in several activity sectors that constrain institutions to be in conformity with them. The obligation to respect certain legal constraints relevant to information technologies security insists on the need for implementing efficient security measures. Mention can be made for examples of Sarbanes-Oxley Act [12], Basel II accord [13], Gramm-Leach-Bliley Act (GLBA) [30], Health Insurance Portability and Accountability Act (HIPAA) [31], EU's Privacy and Electronic Communications (Directive 2002/58) [32], the Data Protection Directive (Directive 95/46/EC)[33], and EuroSox [34].

Confidentiality and integrity of data is most often required by these regulations without specifying the kind of technology to be used. Adopting Quantum Cryptography can be a solution to certify that information is correctly protected and satisfy legal compliance to "*protect informational assets in the best way* [12].

### 6.3 Having a long term vision and being effective

Using quantum cryptography gives the opportunity to invest into a technology with a high potential of remaining secure for a long time.

Working with a security long run vision places an organization in a higher competitive stage than others.

Quantum cryptography contributes to giving enterprises and organisations a real **competitive advantage**.

Knowing that quantum cryptography is reliable allows management attention to be focused on other problems.

Quantum cryptography related costs can be balanced with security efficiency in the long term.

Quantum cryptography contributes to fulfil the objective to have better security without having a great amount of added costs.

### 6.4 Being now at a better security level

Being able to conduct business with a much higher level of confidentiality than the competitors, will lead to several advantages generating direct and indirect profit [35].

The use of an inherently secure way to communicate should state that the organization using quantum cryptography is not anymore at the same level of insecurity as its competitors [6].

### 6.5 Avoiding a security gap

A digital security gap will be amplified between those who use quantum cryptography and those who do not.

### 6.6 Improving some limitations

Being a technology, quantum cryptography naturally presents some limits which are subject to an ongoing process of improvement.

In the current stage, the low-key exchange rates constitute a limit. Albeit current rates in the range of kilobits per second already permit to be operational, the performance of single QKD links increases. The combination of QKD with classical symmetric cryptography helps in this direction. Short distances are also a concern today but the distances are constantly growing and as stated above the system is already operational thanks to the use of the network approach with trusted repeater stations. This is a very useful approach for metropolitan area size networks, which would be the initial applications of quantum cryptography [28].

The implementation of a quantum cryptography network is expensive, like any other innovation's implementation, but it does not consume disruption costs as it can be deployed in parallel to existing key distribution channels. Actually, the quantum cryptography network implementation costs are principally concentrated on hardware or device related costs, which are not as expensive as administrative costs of a service disruption or a service upgrading to another technology. In all the cases, a quantum cryptography network offers a long run service with dedicated implementation costs being easily redeemable [25].

QKD network implementation needs dedicated dark fibre currently. With up to 1000 strands in contemporary fibres, dedicated fibres are not a very costly problem. In spite of this, the future development could allow the use of wavelength-division multiplexing (WDM). It is a technology to multiplex multiple optical carrier signals on a single optical fibre by using different wavelengths (colours) of laser light to carry different signals. This allows a multiplication in capacity, in addition of making it possible to perform bidirectional communication over one strand of fibre.

Apart from the mentioned advantageous features, quantum key distribution (QKD) links have some limitations: the distance over which keys can be exchanged is roughly limited by 100 kilometres today and potentially up to 200 kilometres in the future. The key generation rate exponentially decreases with distance and is currently limited to several tens of Kilohertz. Also is the direct point-to-point nature of QKD links susceptible to denial of service attacks, once the connecting optical fibre is severed. To overcome the previously mentioned limitations in distance, rate, and availability, most recent developments go in the direction **of building redundant networks** out of QKD links. These can be realized either as '**full quantum network**', where quantum signals are refreshed in the network nodes using so-called quantum repeaters [36] [37] or as 'classical trusted relaying network'. Quantum repeaters however are merely theoretical constructs by now as they require elaborated quantum operations and especially quantum memories which are neither available today, nor are expected to be available in a foreseeable future. Trusted relay networks, on the other hand, can be implemented with today's technologies. They follow a simple principle: local keys are generated over QKD links and then stored in the nodes on both ends of the link – the trusted relays. The key distribution between distant (not directly connected) nodes is performed over a QKD path, i.e. a one-dimensional chain of trusted relays connected by QKD links. Secret keys are forwarded, in a hop-by-hop fashion, along the QKD path. End-to-end information theoretic security is thus obtained between the end nodes, provided that the intermediary nodes can be trusted. Trusted relay networks can be used to build wide area QKD networks – but the assumption of trusted nodes has certain implications on how these networks can be used for electronic commerce. This network paradigm with multiple links originating from one network node can overcome distance limitations of QKD, as well as capacity and availability limitations through redundant paths [28].

### 6.7 Cost model

Proposing a real cost model for network architecture solutions is very difficult because each case is particular and has to answer a special well identified need. Quantum cryptography is nowadays affordably compatible to the requirement of high-speed networks and of high security level.

Is quantum cryptography expensive?

It depends on what "expensive" does mean in a particular context. Currently adopting a Quantum Key Distribution (QKD) technology has a good Return on Investment (RoI) for many reasons:

- QKD will remain secure for a long time because of its physical properties;

- there is no need for frequent upgrades which is a very expensive aspect in a long

run vision;

- there is no disruption of business during upgrade, which is more expensive than QKD by itself.

The implementation of a new and innovative technology could generate costs due principally to hardware costs. But it is not to be neglected that costs can be reduced in the long term by massive market adoption.

Some consider Quantum Cryptography as approximately twice as expensive as classical cryptography. Most products are attractive in terms of costs because of scalability properties and effectiveness. Adopting quantum technologies to secure point-to-point connections between two distributed buildings is possible in a range of cost comparable to other existing solutions[4].

### 6.8 A market and industrial perspective

Quantum cryptography has great potential to become a key technology for securing communication confidentiality and privacy in the future information society and thus to become a driver for the success of a series of services in the field of e-government, e-commerce, and e-health [38]. Several studies have already demonstrated the need to have pro-active security and efficient security countermeasure [15, 39] to face the challenges linked to ICT misuses and cybercrime issues [40-43] because quantum technology has the potential to break the vicious circle between the fight between code makers and code breakers.

A quantum industry has recently emerged [44] and reliable actors already exist and propose technologies and services.

"The USA Department of Defense (DoD) currently funds several quantum-cryptography projects as part of a $20.6 million initiative in quantum information. Globally, public and private sources will fund about $50 million in quantum-cryptography work over the next several years. Andrew Hammond, a vice president of MagiQ, estimates that the market for QKD systems will reach $200 million within a few years, and one day could hit $1 billion annually"[5].

### 6.9 Social impacts

The use of quantum key distribution for robust secret transmission will change the actual network security paradigm. This will affect without precedence the way to build trust among distributed actors.

---

[4] For example : A typical point-to-point link configuration (considering a Quantum Key Distribution device and the link) costs about 65'000 €. Securing two remote points will cost about 130'000 €. This price is subject to decrease with the number of links to be connected and it is subject to change from one quantum technology provider to another. Source : IdQuantique products (www.idquantique.com/).

[5] Source : http://www.aip.org/tip/INPHFA/vol-10/iss-6/p22.html American Institute of Physics – December 2004/January 2005

The information society, as the development of an efficient digital economy, could not exist without strong security mechanisms. Quantum security technology can contribute to build confidence in the use of ICT's. It will enforce the consumers' willingness to use electronic services and develop economic activities for the benefit of all [45].

More and more e-voting applications are under development worldwide. Some initiatives have been already implemented and are in use.

The citizen e-voting information integrity cannot be guaranteed without integration of at least the pure random number generation possible with quantum technology to generate cryptographic keys for the e-voting process. In this particular e-government context, achieving strong security through efficient key generation and distribution is mandatory to support the democratic process[6].

# 7 STANDARDISATION OF QUANTUM CRYPTOGRAPHY AND QUANTUM TECHNOLOGIES

## 7.1 Standardisation in SECOQC: An ETSI answer

In the course of the SECOQC project, the need for standardisation arose only gradually when we were confronted with the need to combine five technologically different quantum cryptographic key distribution link technologies in the SECOQC Quantum Back Bone network. A common interface was developed and published as **open standard.** Documentation and software emulator are available on the www.secoqc.net homepage. In addition, standardised components and assumptions are indispensable for the definition and development of effective and reliable security metrics for the security evaluation of quantum cryptographic systems [28].

Together with the European Telecommunications Standards Institute ETSI, the University of Lausanne in close collaboration with Austrian Research Centers ARC developed the plan to initiate an Industry Specification Group (ISG) [46].

The ISG[7] on Quantum Cryptography and Quantum Technologies has been founded on July 2008 to bring together relevant actors from science, industry, and commerce from Europe and from abroad. Among the initial consortium of signatories are the Austrian Research Centers ARC, Telefonica S.A., Polytechnical University Madrid, TELECOM Paris Tech, Hewlett-Packard, Toshiba Research Europe, idQuantique S.A., SmartQuantum, Quinetiq Ventures, University of Lausanne, and with pending application process (by the time of Sept 2008): Thales S.A., European Space Agency, Swisscom, Mimas Berhad, Istituto Nazionale di Ricerca Metrologica.

Overall goal of the ETSI – ISG initiative is to offer an open forum with significant leverage effects on coordination, cooperation and convergence. Communication between producers, developers, and scientists on the one side, and prospective customers and users on the other side shall help to promote mutual understanding. To achieve these goals, two separate tracks, one for business application and one for technical standards will be initiated.

---

[6] http://www.ge.ch/evoting/ (online 30.09.2008)
[7] http://www.etsi.org/WebSite/NewsandEvents/ISG_QKD.aspx (online 30.09.2008)

Standardisation activities can support the commercialisation of quantum cryptography on various levels and stages. Gaps on requirements, technology, and application level, which are still prevalent in today's quantum cryptographic systems shall be narrowed and eventually closed.

Business standards deal with standardisation issues that are not entirely focused on technology. They are concerned with the analysis of business requirements for different groups of prospective users, as well as with the development and definition of suitable interfaces to guarantee connectivity to existing infrastructures and compatibility to existing service management [47]. Of specific interest is the application interface over which key requests are issued to a quantum cryptographic key distribution link, or network, and over which subsequently the generated and distributed keys are handed back to the application. These issues will be addressed by a technical standardization process.

### 7.2 Standards to fulfil business requirements

There are different groups of prospective users for quantum key distribution systems, who all may impose different requirements on these systems. These requirements are in most cases imposed by compulsory **security policies** that reflect the respective security need for such classes of users. Example user groups are banks, governmental institutions, and health institutions. The ISO/EN 15408 'Common Criteria' standard [48-50] provides a suitable formalism, the **Protection Profile**, to lay down implementation independent security objectives for quantum cryptographic systems as they are required from a business point of view.

The security objectives defined for quantum key distribution systems in the Technical Standards track must match the **security objectives** of the requirements for a certain user group. The use of the same formalism helps to mediate developers and users – i.e. to bring them more into contact so that they learn about their respective technical possibilities and business needs.

### 7.3 Technical Standards for quantum key distribution (QKD)

**Standardized interfaces** are needed in order to allow a quantum cryptographic key exchange system to be attached to existing information communication technology systems. Classical (i.e. non-quantum) key exchange systems are available on the market and are widely used to exchange keys for securing data transfer.

These systems do in general employ Diffie-Hellman style asymmetrical key exchange. Quantum cryptographic systems replacing such asymmetrical key exchange subsystems shall be compatible to these interfaces. In course of this task, a collection of interfaces relevant for quantum cryptographic systems shall be compiled, and the feasibility of adapting interfaces to specific characteristics of quantum key exchange shall be analysed .The main goal of this track is to progress towards technical standards for quantum key distribution systems.

One task is the development of quantum key distribution systems specifications 'from bottom up', i.e. from what is already realised in contemporarily available quantum key distribution links and networks. Furthermore, properties of specific macroscopic quantum optical components, like photon sources and detectors, and interfaces between components of quantum cryptographic systems shall be subject to standardisation. This shall facilitate better compatibility between components of different origin in single quantum key distribution links.

Quantum key distribution links usually consist of single modular components, which can be clearly distinguished from other components of the system. Examples for such components are: photon sources, photon detectors, or the computers or embedded systems, which perform the key distillation. For these components it is useful to define standardised security properties, like e.g. multi photon pulse probabilities for single photon sources.

In addition, it is necessary to agree on common interfaces for modular components, in order to facilitate the integration of components of different vendors. This activity shall be complementary to the standardisation of components as regarding to physical and security related properties. Together, these measures shall reduce development effort for quantum cryptographic systems significantly.

### 7.4 Quantum key distribution security specification

The ISO/EN 15408 'Common Criteria' standard can also be used for technically detailed **quantum key distribution security specifications**. They contain a **threat and risk analysis** for the assets that are to be protected in the system (the produced keys). Based upon this analysis, a number of security objectives shall be derived, which again are to be maintained during operation of the quantum cryptographic system. Consequently, specific functional requirements for actual implementations of quantum cryptographic systems shall be developed and listed. These low level specifications shall provide guidance for developers and manufacturers of quantum cryptographic systems who may develop and produce their systems in accordance to these specifications.

At a later instant of time, security evaluations by accredited evaluation laboratories may be conducted to prove the concordance between security specifications and implemented systems (accreditation).

In the ontology of the 'Common Criteria' such implementation dependent security specifications are so-called '**Security Targets**' – in contrast to the above mentioned implementation independent Protection Profiles. Ideally the Security Targets (bottom-up specifications – '**what we can deliver**') meet with the Protection Profiles (top-down specifications – '**what we need**'). Thus it shall be ensured that a quantum cryptographic system delivers exactly what an interest group requires.

## 8 CONCLUSION

**Achieving confidentiality** is one of the cornerstones of security measures, which could be fulfilled by cryptographic implementations. The reliability and robustness of the cryptographic mechanisms essentially rely upon cryptographic keys (key generation, distribution and storage, key secrecy). With the increase of computational power contemporary encryption and decryption methods based on secret keys for securing communication are **under threat**. Information becomes a very important asset for today's organizations, which are more and more subject to regulatory compliance issues. Added to the fact that information security officers could be accountable for a lack of sufficient information security measures (civil and penal responsibilities) they have **to rely upon strong technical security solutions**. The security problems related to classical encryption methods may no longer satisfy the requirement for a strong security level. Quantum cryptography contributes to answering these needs.

Cryptographic solutions must support reliable and provable confidentiality services to support today's business competition and effectiveness in an uncertain world. It has been demonstrated that if underling cryptographic mechanisms are based on computational assumptions, they are no longer sufficiently secure. The only possibility to bypass this fact is to change the cryptographic paradigms by integrating quantum theory into cryptographic solutions to exploit its inherent secure mechanisms.

Rethinking fundamentals in cryptography is the only solution to develop a new vision of security for the benefit of transactions that are critical for institutions and people.

This allows introducing a break into the vicious circle, which assumes that only the entity that offers commercial security solutions can master institutions' data confidentiality. With quantum key distribution**, institutions and people have for the first time the means to be sure that their data are under their own control and cannot be obtained by eavesdroppers without knowledge of sender or recipient**.

Adopting quantum random number generators can be a first step to enforce actual cryptographic robustness in every day transactions. This can be done very easily and is cost effective. The second move towards high security is **to transmit confidential data trough point-to-point connections secured by the combination of quantum key distribution and strong classical encryption algorithms**. This choice has already been made for high value applications and long term secure data retention by leading institutions that are security aware in order to obtain a strong competitive advantages in the market.

The last step towards **a truly secure network environment** will be realised when organisation will design their network architecture based upon quantum cryptographic network solutions. The feasibility of this approach has been demonstrated in October 2008 by the SECOQC consortium in an experimental deployment of a quantum cryptographic network with several nodes and links in Vienna. This experiment demonstrates a realistic prototype contributing to conceive innovative networks that integrate an intrinsic confidence level.

The SECOQC quantum cryptographic network shows that transferring quantum key distribution from the controlled and well-defined laboratory environment into **a real-world commercial environment** is possible. This concerns the development of networked structures, which better meet the requirements of different prospective user groups, as well as the development of standardised procedures for generating and enabling trust into quantum cryptography systems.

## 9 FREQUENTLY ASKED QUESTIONS ABOUT QUANTUM CRYPTOGRAPHY

### 9.1 Do I need quantum cryptography?

That depends on the importance given to informational assets. Two important components have to be considered:

- The security level the currently mechanism offer,
- The value of the information to be protected and security criteria to be fulfilled.

In some critical infrastructures like governmental and administrative applications, financial domains, energy sector installations etc. the need of a robust technology offering a high security level is something mandatory.

Taking into account that computer power drastically increases in a continuous way and that there is no mathematical proof of its cryptographic algorithms' robustness, asymmetrical cryptography could be considered as potentially insecure in the near future.

### 9.2 Do international regulations require the use of quantum cryptography?

No regulation requires explicitly the use of quantum cryptography but regulations (such as SoX and Basel II) regarding the financial domain require exactitude and reliability for the financial statement. However what is required is to protect informational assets in the best way, which could be differently interpreted according to the context. Quantum cryptography is a solution satisfying legal compliance and certifying that information is correctly protected.

### 9.3 Is QKD expensive?

It depends on what "expensive" does mean in a particular context. Currently, adopting a QKD technology has a good Return On Investment for many reasons:

- QKD will remain secure for a long time because of its physical properties;
- there is no need for frequent upgrades which is a very expensive aspect in a long run vision;
- there is no disruption of business during upgrade, which is more expensive than QKD itself.

The implementation of a new and innovative technology could generate cost due principally to hardware costs.

But it is not to be neglected that cost can be reduced in the long term by massive market adoption.

### 9.4 Why do I need to change my present cryptographic system?

First of all you do not need to change your present cryptographic system. QKD is for generating and distributing secure keys for your existing cryptographic applications, but also for your future applications. In that way the adoption of a QKD will improve significantly the quality and the security level of communication.

### 9.5 How to integrate quantum cryptography in my existing IT infrastructure?

The high speed QKD system can be integrated into a fibre optical telecom infrastructure, such as a Local Area Networks (LAN) or Metropolitan Area Networks (MAN) to enable secure communication and information exchange among users.

It could be installed in parallel to the running system, and then eventually been plugged in. Another way is to wrap the QKD system around your existing systems.

### 9.6 Does QKD need dedicated optical fibre?

QKD's network implementation needs dedicated fibre currently. With up to 1000 strands in contemporary fibres, dedicated fibres are not a very costly problem. In spite of this, the future development could allow the use of wavelength-division multiplexing (WDM). It is a technology, which multiplexes multiple optical carrier signals on a single optical fibre by using different wavelengths (colours) of laser light to carry different signals. This allows a multiplication in capacity, in addition to enabling bidirectional communications over one strand of fibre.

Alternatively, free space QKD can be applied where photons are exchanged between two telescopes (see below).

### 9.7 Is it possible to use QKD on a copper fibre?

QKD explicitly cannot be use on copper fibre, but the use of a quantum number generator to design a key session (as described in the e-voting application of the Geneva Canton - CH) is possible over a copper fibre.

### 9.8 Is QKD possible in free space?

The possibility to operate QKD in free space was demonstrated on multiple occasions (1998: 1 km in the US, 2001: 1.9 km in the UK, 2002: 10 km in the USA). In 2003 photons were exchanged in the Alps in Germany at a distance of 23 km. In 2005 it was reported that Chinese scientists succeeded in an inner-city free-space distribution of entangled photon pairs over a distance of 10.5 km. About the same time, an experiment was published where entangled photons were distributed directly through the atmosphere to a receiver station in 7.8 km distance in the City of Vienna. In 2007 a quantum key was established between ESA-telescopes on two of the Canary Islands over a distance of 144 km. The quality of the free space transmission is highly depending on atmospheric conditions.

### 9.9 Is it possible to use QKD within a Local Area Network (LAN)?

A local area network implementation of QKD is already operational to connect one or more remote sites by a QKD link. The transmission distance can be increased beyond 100 km by chaining links.

### 9.10 What is the distance limitation?

For a point-to-point connection the distance limitation is about 100 km in order to obtain reliable quality of the signal. The SECOQC's network architecture allows to overpass this physical constraint and to extend the capacity of QKD transmission over a wide area network (WAN).

### 9.11 Which are the principal domains in which a QKD network might be useful?

Quantum key generation and distribution answer high security needs and can for example be used to secure transmissions in the following areas:

- Government or military entities
- Banks and financial institutions (Interbank-Transfer, communications Bank to ATM,…);
- Critical infrastructures,
- Telecoms operators;
- Airports;
- Etc.

### 9.12 Can end users benefit from quantum cryptography?

Any end user disposing of an optical fibre infrastructure can benefit from the Quantum Cryptography advantages to enhance the security level of their transmission.